\documentclass[sigconf]{acmart}
\AtBeginDocument{%
  }
\usepackage[linesnumbered,vlined,ruled]{algorithm2e}
\usepackage{multirow}
\usepackage[most]{tcolorbox}
\usepackage{graphicx}
\usepackage{subcaption}
\usepackage{float}

\setcopyright{none}

\begin{document}

\title{Agentic Mixed-Source Multi-Modal Misinformation Detection with Adaptive Test-Time Scaling}

\author{Wei Jiang}
\affiliation{%
  \institution{The University of Queensland}
  \city{Brisbane}
  \country{Australia}
}
\email{wei.jiang@uq.edu.au}

\author{Tong Chen}
\affiliation{%
  \institution{The University of Queensland}
  \city{Brisbane}
  \country{Australia}}
\email{tong.chen@uq.edu.au}

\author{Wei Yuan}
\affiliation{%
  \institution{The University of Queensland}
  \city{Brisbane}
  \country{Australia}}
\email{w.yuan@uq.edu.au}

\author{Quoc Viet Hung Nguyen}
\affiliation{%
  \institution{Griffith University}
  \city{Gold Coast}
  \country{Australia}}
\email{henry.nguyen@griffith.edu.au}

\author{Hongzhi Yin}
\authornote{Corresponding author.}
\affiliation{%
  \institution{The University of Queensland}
  \city{Brisbane}
  \country{Australia}}
\email{h.yin1@uq.edu.au}

\renewcommand{\shortauthors}{Wei Jiang et al.}
\begin{abstract}
Vision-language models (VLMs) have been proven effective for detecting multi-modal misinformation on social platforms, especially in zero-shot settings with unavailable or delayed annotations. However, a single VLM's capacity falls short in the more complex mixed-source multi-modal misinformation detection (M$^3$D) task. Taking captioned images as an example, in M$^3$D, the false information can be sourced from either untruthful texts, forged images, or a mismatch between two modalities. 
Although recent agentic systems can handle zero-shot M$^3$D by connecting modality-specific VLM agents, their effectiveness is still bottlenecked by their agentic architecture. In existing agentic M$^3$D solutions, for any input sample, each agent only performs one-time forward reasoning, making its decision prone to inherent model randomness and reasoning errors in challenging cases. Furthermore, the lack of exploration on alternative reasoning paths means that the reasoning capacity of modern VLMs is not fully capitalized. In this work, we present AgentM$^3$D, a multi-agent framework for zero-shot M$^3$D. To amplify VLMs' reasoning capability, an innovative adaptive test-time scaling paradigm is proposed, where AgentM$^3$D scales the reasoning of each modality-specific VLM agent via the best-of-$N$ mechanism, and a critic agent is designed to achieve task-aligned scoring. All agents are placed in a cascading, modality-specific decision chain, so as to reduce unnecessary agent use and limit error propagation. In addition, to keep AgentM$^3$D scalable, a planning agent dynamically determines the maximum number of reasoning paths needed based on sample difficulty, and an adaptive stopping mechanism is in place to further prevent each agent from overthinking. Extensive experiments on two M$^3$D benchmarks show that AgentM$^3$D achieves state-of-the-art zero-shot detection performance compared to various VLM-based and agentic baselines. 
\end{abstract}





%

\maketitle


\section{Introduction}

Multi-modal misinformation on social media evolves rapidly, with new manipulation patterns continuously emerging~\cite{o2019misinformation}, including fake news, image forgery, narrative edits, and so on. In such dynamic environments, collecting high-quality labeled data that covers newly appeared forgery types is often infeasible or severely delayed~\cite{singh2020first,pennycook2019fighting}. As a result, supervised multi-modal misinformation detection (MMD) systems struggle to remain effective in real-world deployment. 
This makes zero-shot MMD a practical and timely demand, and there has been an uptake~\cite{lin2023zero,abdali2024multi} on developing such solutions.

\begin{figure}[t]
    \centering
    \includegraphics[width=.9\linewidth]{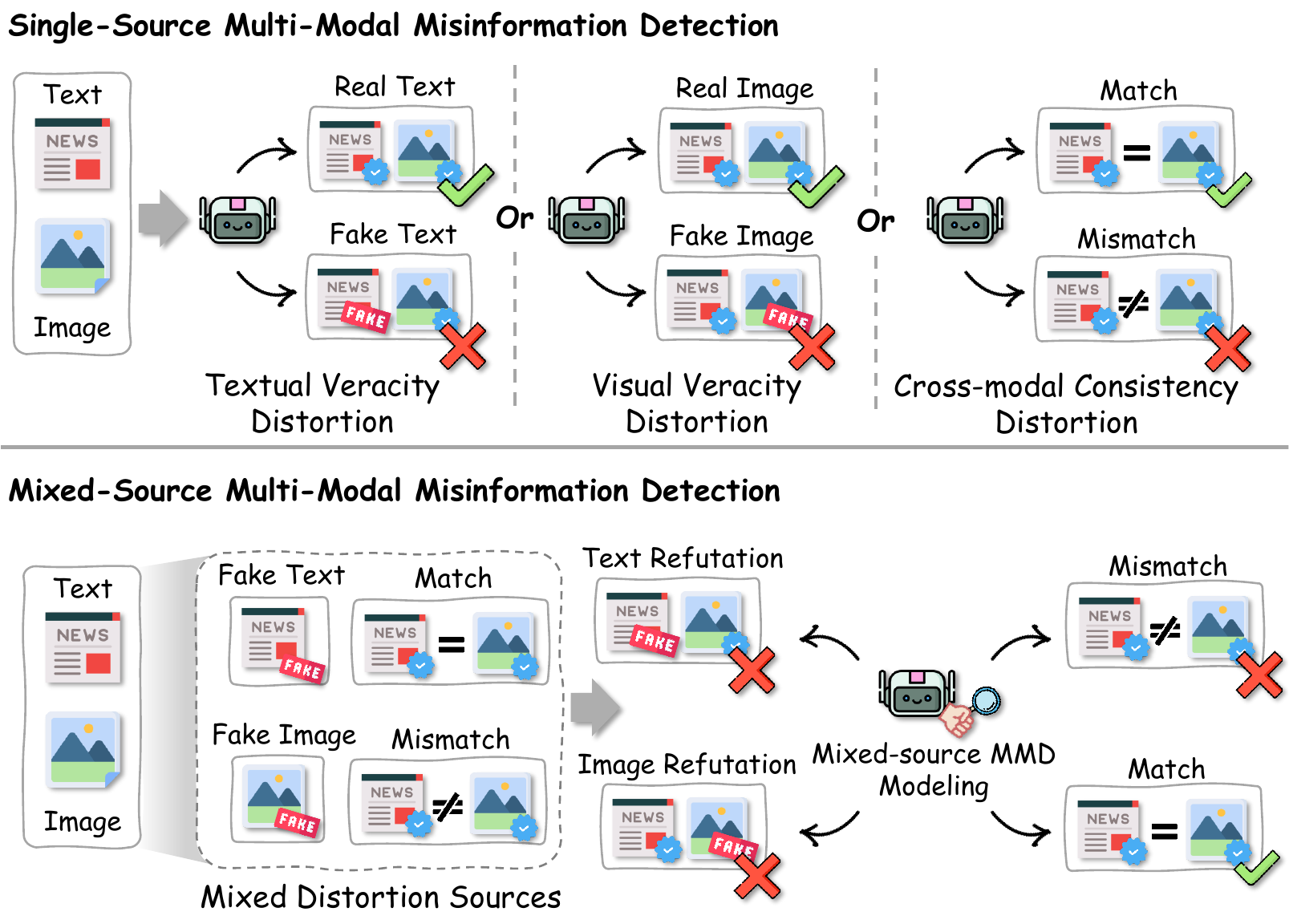}
    \caption{Comparison between single-source and mixed-source multi-modal misinformation detection.}
    \label{fig:agentM3D_intro}
\vspace{-.7cm}
\end{figure}

Recently, vision-language models (VLMs)~\cite{zhang2024vision}, pretrained on large-scale image-text corpora, have exhibited strong cross-modal reasoning and generalization capabilities, and are therefore widely adopted for zero-shot multi-modal misinformation detection~\cite{liu2024mmfakebench,shang2024mmadapt}.
However, most existing VLM-based MMD approaches implicitly operate under simplified assumptions about misinformation sources. In particular, they primarily focus on scenarios where misinformation shares one same manipulation pattern, overlooking the fact that real-world misinformation often exhibits mixed-source characteristics, as illustrated in Fig.~\ref{fig:agentM3D_intro}. In the mixed-source multi-modal misinformation detection (M$^3$D) setting, false information may originate from various manipulation sources. As per Fig.~\ref{fig:agentM3D_intro}, in news with images, misinformation can be sourced from untruthful textual claims, forged or manipulated images, or subtle mismatches between otherwise authentic modalities~\cite{liu2024mmfakebench}. This practical setting further extends the detection task from binary classification to multi-class classification \cite{t2agent,liu2024mmfakebench}, so as to provide precise attribution to the exact source of counterfeit. Consequently, the strong interdependencies among these sources makes M$^3$D fundamentally challenging for single VLM-based solutions. Specifically, M$^3$D demands stage-wise conditional reasoning: textual claims should be assessed independently of visual content, visual evidence must be verified against objective facts, and cross-modal alignment should be evaluated under the outcomes of prior decisions. When a single VLM is forced to reason over all these coupled aspects simultaneously, even minor inaccuracies in one dimension can propagate and lead to incorrect final predictions~\cite{zhou2024finefake,shen2024small}.

As a remedy, recent work has explored agentic frameworks for zero-shot M$^3$D  that decompose detection into modality-specific reasoning stages~\cite{liu2024mmfakebench}. By connecting multiple specialized VLM agents, these systems introduce structured inference and partially reduce decision coupling compared to monolithic models. Nevertheless, existing agentic M$^3$D approaches remain fundamentally constrained by their inference paradigm. For each input sample, individual agents typically perform a single forward reasoning pass, rendering their predictions vulnerable to inherent model stochasticity and brittle reasoning in challenging cases. More critically, the lack of systematic exploration over alternative reasoning paths prevents modern VLMs from fully leveraging their latent reasoning capacity, particularly in complex mixed-source scenarios where a single reasoning trace is often insufficient.
Some recent agentic systems attempt to alleviate this limitation by introducing planning modules and extensive tool invocation~\cite{t2agent}. For instance, T$^2$Agent employs a planner to decompose each input into multiple tool-solvable subtasks and prioritize different information sources. However, these subtasks are still executed via single-pass VLM reasoning, leaving the fundamental inference bottleneck of one-shot decision making largely unaddressed. Moreover, such designs rely heavily on third-party components, whose availability and reliability are often uncertain, potentially compromising system robustness in practical deployment.
Consequently, despite recent progress, building effective zero-shot VLM-based systems for M$^3$D remains an open challenge.

In this work, we seek to improve the effectiveness of M$^3$D systems by taking full advantage of the reasoning capacity of VLMs at inference time through test-time scaling (TTS) \cite{chen2024expanding}. TTS enhances inference reliability by exploring and evaluating multiple reasoning trajectories without additional training, and has proven effective for strengthening large models under challenging decision settings~\cite{zhang2025survey}.
However, directly applying TTS to M$^3$D introduces several non-trivial challenges. 
First, \emph{when to scale}: the reasoning difficulty varies substantially across samples, and some easy cases can be accurately resolved with a single forward pass. Uniformly applying test-time scaling to all inputs therefore incurs unnecessary computational overhead without clear benefits.
Second, \emph{how much to scale}: even for samples that require enhanced reasoning, the difficulty differs across misinformation sources and modalities. Over-scaling simple reasoning paths not only increases inference cost but may also lead to overthinking and degraded decision quality, making principled stopping criteria essential.
Third, \emph{how to score reliably}: test-time scaling relies on reward models to evaluate multiple reasoning trajectories, yet general-purpose reward models are not aligned with the complex decision requirements of M$^3$D and may misjudge reasoning quality in mixed-source settings. To date, no existing framework has jointly addressed these challenges in a scalable and cost-effective manner for zero-shot M$^3$D.

To address these challenges, we propose \textbf{AgentM$^3$D}, a multi-agent framework that leverages adaptive test-time scaling to enhance the intrinsic reasoning capability of VLMs for zero-shot mixed-source multi-modal misinformation detection. AgentM$^3$D organizes modality-specific detection agents in a hierarchical cascade, mitigating decision coupling and limiting error propagation across inference stages. 
Specifically, to determine \emph{when} enhanced reasoning is required, AgentM$^3$D introduces a lightweight planning agent that selectively activates test-time scaling based on sample difficulty. To ensure \emph{reliable scoring} of scaled reasoning candidates, the framework integrates critique-aware Best-of-$N$ reasoning~\cite{sessa2407bond}, in which modality-specific critique signals complement general-purpose reward models to provide task-aligned evaluation. To regulate \emph{how much} computation is warranted, an adaptive early-stopping mechanism dynamically controls the extent of reasoning exploration. Together, these components transform test-time scaling from a brute-force heuristic into a structured, decision-aware inference process for robust and cost-efficient zero-shot M$^3$D.

Our contributions are threefold:
\begin{itemize}
    \item To the best of our knowledge, we are the first to provide a systematic characterization of the key challenges of applying TTS in M$^3$D, including when to scale, how to assign reliable scores, and how much computation is warranted.
    \item We propose AgentM$^3$D, a multi-agent system with adaptive test-time scaling that jointly addresses when to scale, how to score reliably, and how much computation is warranted in zero-shot settings.
    \item We empirically validate AgentM$^3$D on multiple multi-modal misinformation benchmarks, demonstrating improved effectiveness and test-time efficiency over strong VLM-based and agentic baselines.
\end{itemize}

\begin{figure*}
    \centering
    \includegraphics[width=.83\linewidth]{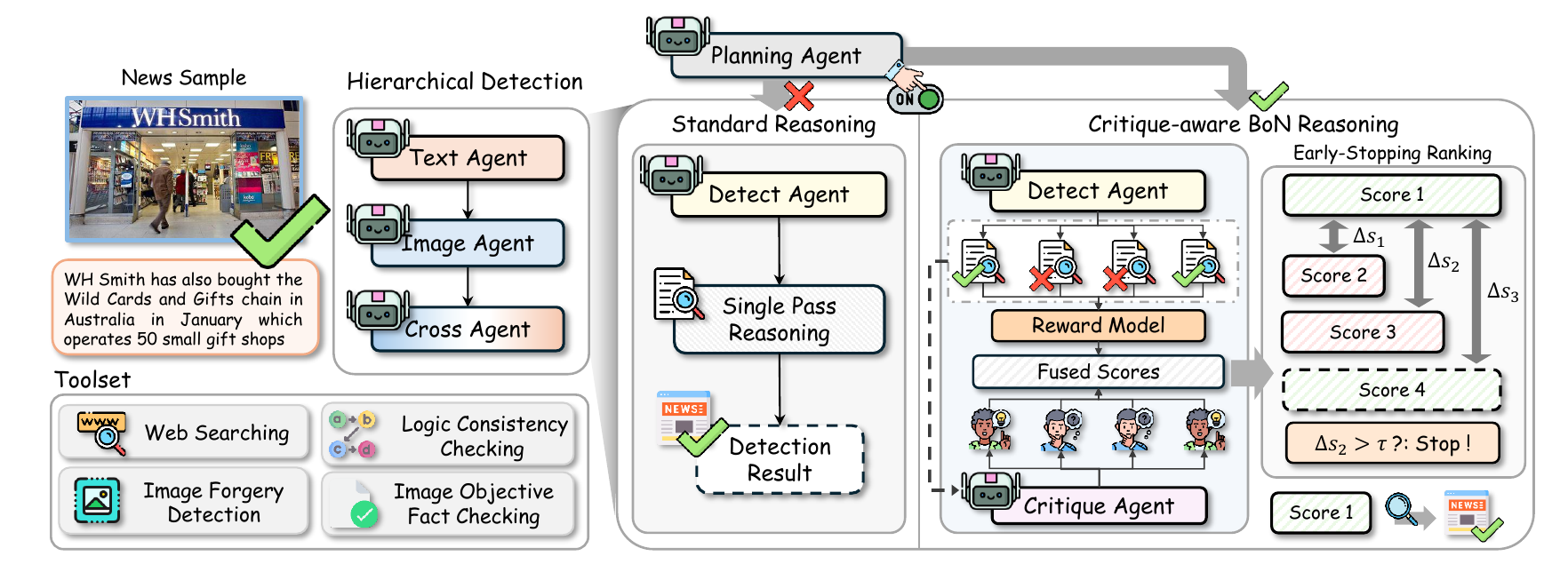}
    \caption{The overall structure of AgentM$^3$D. A planning agent routes each input to either standard reasoning or critique-aware Best-of-$N$ reasoning. The latter explores multiple reasoning trajectories, integrates reward and critique signals for candidate selection, and applies adaptive early-stopping ranking to dynamically terminate inference when confident scores emerge.}
    \label{fig:agentM3D_overview}
\end{figure*}

\section{Preliminaries}
\noindent {\textbf{Problem Formulation.}}
We study the problem of mixed-source multi-modal misinformation detection (M$^3$D). Formally, a collection of news is represented as
$\mathcal{D}_{\text{news}} = \{(T_i, V_i, y_i)\}_{i=1}^{N}$,
where each item consists of a textual claim $T_i$, an associated news image $V_i$, and a ground-truth label $y_i \in \{0,1,2,3\}$.
The label space corresponds to four mutually exclusive categories:
original (real), textual veracity distortion, visual veracity distortion, and cross-modal consistency distortion, respectively. 

We formulate M$^3$D as a zero-shot multi-class classification task. Given an input sample $(T_i,V_i)\in\mathcal{D}_{\text{news}}$,
a single agent $a$ produces a categorical prediction
along with an associated natural language explanation:
\begin{equation}
(y_{i,a}, r_{i,a}) \sim \phi_a(T_i,V_i,\mathcal{P}_a),
\end{equation}
where $r_{i,a} \in \mathcal{R}$ is the corresponding explanation in natural language and $\mathcal{P}_a$ is the prompt of agent $a$.

We further extend this formulation to a multi-agent system $\mathcal{A}=\{a_1,a_2,\ldots,a_m\}$, in which agents may assume different roles and operate collaboratively. Agents interact by incorporating the outputs of other agents as additional inputs. If agent $a_j$ is associated with a set of agents $\mathcal{N}(j)\subseteq\mathcal{A}$, then its inference depends on both the input $(T_i,V_i)$ and the outputs of all related agents: 
\begin{equation}
(y_{i,a_j}, r_{i,a_j}) \sim
\phi_{a_j}\!\left(T_i,V_i,\{(y_{i,a_k}, r_{i,a_k}) \mid a_k \in \mathcal{N}(j)\} , \mathcal{P}_{a_j}\right).
\end{equation}

\noindent {\textbf{Test-Time Scaling via Best-of-$N$ (BoN).}}
Beyond single-pass inference, test-time scaling (TTS) strategy can be considered to enhance the reasoning robustness of individual agents. Specifically, given an input sample $(T_i,V_i)$ and an agent $a$,
it allows the agent to perform multiple independent inference trials under the same prompt $\mathcal{P}_a$. Formally, the agent is stimulated $N$ times to produce a set of candidate outputs:
\begin{equation}
\{(y_{i,a}^{(m)}, r_{i,a}^{(m)})\}_{n=1}^{M}
\sim \phi_a(T_i,V_i, \mathcal{P}_a),
\end{equation}
where each trial corresponds to an independent stochastic realization of the inference process of the agent.

The BoN strategy selects a single output from these candidates according to a predefined selection criterion $\mathcal{S}$:
\begin{equation}
(y_{i,a}^{*}, r_{i,a}^{*})
=
\mathcal{S}\!\left(\{(y_{i,a}^{(m)}, r_{i,a}^{(m)})\}_{n=1}^{M}\right).
\end{equation}

Intuitively, BoN serves as a form of TTS, allowing the agent to explore diverse reasoning paths and retain the most reliable prediction, thereby improving robustness without additional training.


\section{Methodology}
In this section, we present the technical details of AgentM$^3$D, a multi-agent framework for mixed-source multi-modal misinformation detection. AgentM$^3$D adopts a hierarchical and adaptive design, in which specialized agents are sequentially invoked to assess textual veracity, visual authenticity, and cross-modal consistency in a structured manner. We first introduce the hierarchical modality-specific detection agents in Sec.~\ref{sec:detect_agent}, which decompose misinformation detection into a cascade of text-, image-, and cross-modal inference stages under one-time forward reasoning. While effective, this inference paradigm is inherently sensitive to stochastic reasoning variability and may underexploit the full reasoning capacity of modern VLMs.
To address this limitation, Sec.~\ref{sec:critque_agent} presents a test-time scaling mechanism based on selective Best-of-$N$ (BoN) reasoning with critique-aware ranking, enabling agents to explore multiple reasoning trajectories and improve robustness. Finally, to balance detection performance with inference efficiency, Sec.~\ref{sec:planning_agent} introduces a lightweight planning strategy to dynamically determine when test-time computation should be allocated. The overall structure of AgentM$^3$D is illustrated in Fig.~\ref{fig:agentM3D_overview}.

\subsection{Hierarchical Modality-specific Detection}
\label{sec:detect_agent}

Inspired by MMD-agent~\cite{liu2024mmfakebench}, AgentM$^3$D decomposes multi-modal misinformation analysis into hierarchical reasoning stages. However, in MMD-Agent, this decomposition is realized through fixed prompt chaining within a single vision--language model, where intermediate reasoning results are implicitly embedded into subsequent prompts.
In contrast, AgentM$^3$D explicitly instantiates each reasoning stage as an independent detection agent with a well-defined input--output interface. 
The resulting agents are organized into a hierarchical cascade with conditional activation, which progressively refines the detection decision from unimodal veracity assessment to cross-modal consistency checking. In particular, the cascade begins with textual veracity detection, as textual claims typically provide the primary semantic signal and can be assessed efficiently with relatively low computational cost, allowing early filtering of distorted content.
Each detection agent produces not only a categorical decision but also an explicit natural-language reasoning trace as part of its output. This explicit agent abstraction decouples stage-wise reasoning from prompt engineering and enables principled modeling of inter-agent dependencies and conditional activation directly at the level of agent inference.

Specifically, AgentM$^3$D employs three kinds of agents to detect textual, visual, and cross-modal misinformation.  

\noindent \textbf{Textual Veracity Detection Agent.}
The textual veracity detection agent is responsible for assessing whether the textual claim $T$ contradicts credible objective evidence. Its inference process is formulated as
\begin{equation}
(y_{\text{text}},\; r_{\text{text}})
\;\sim\;
\phi_{\text{text}}\!\left(T,\; \mathcal{P}_{\text{text}},\; \mathcal{T}_{\text{text}}\right),
\end{equation}
where $y_{\text{text}}$ denotes the predicted textual veracity label produced by the agent, and $r_{\text{text}}$ is the corresponding explicit natural-language reasoning. $\mathcal{T}_{\text{text}}$ represents auxiliary evidence retrieved from an external textual information tool (details are in the Appendix~\ref{sec:toolset}).

\noindent \textbf{Visual Veracity Detection Agent.}
The visual veracity detection agent evaluates whether the image $V$ contradicts credible objective evidence or violates common-sense constraints. Its inference process is formulated as
\begin{equation}
(y_{\text{image}},\; r_{\text{image}})
\;\sim\;
\phi_{\text{image}}\!\left(V,\; \mathcal{P}_{\text{image}},\; \mathcal{T}_{\text{image}}\right),
\end{equation}
where $y_{\text{image}}$ denotes the predicted visual veracity label, and $r_{\text{image}}$ is the corresponding explicit natural-language reasoning. $\mathcal{T}_{\text{image}}$ represents auxiliary observations obtained from an image analysis tool (details are provided in the Appendix~\ref{sec:toolset}).

\noindent \textbf{Cross-modal Consistency Detection Agent.}
The cross-modal consistency detection agent evaluates whether the textual claim $T$ and the image $V$ are semantically aligned, i.e., whether they refer to the same entities, events, and contexts. Unlike the textual and visual veracity agents, this agent does not aim to verify the factual correctness of each modal information. Instead, it targets deceptive pairings in which each modality may appear individually plausible, yet their combination forms a misleading narrative.
Its inference process is formulated as
\begin{equation}
(y_{\text{cross}},\; r_{\text{cross}})
\;\sim\;
\phi_{\text{cross}}\!\left(T,\; V,\; \mathcal{P}_{\text{cross}},\; \mathcal{T}_{\text{image}}\right),
\end{equation}
where $y_{\text{cross}}$ denotes the predicted cross-modal consistency label, and $r_{\text{cross}}$ is the corresponding explicit natural-language reasoning.
We incorporate $\mathcal{T}_{\text{image}}$, an auxiliary observation produced by an image analysis tool that provides grounded descriptions of the visual content. This grounded visual summary makes the image content explicit and comparable to the textual claim, facilitating reliable alignment assessment. Note that we did not use the textual tool $\mathcal{T}_{\text{text}}$ here because it is for truthfulness verification, which is not required for our cross-modal consistency detection.

\noindent \textbf{Hierarchical Cascade Execution.}
AgentM$^3$D organizes the above three detection agents into a hierarchical cascade with conditional activation. The key idea is to progressively refine the detection decision from unimodal veracity assessment to cross-modal consistency checking, while avoiding unnecessary computation.
We first define a distortion indicator function as:
\begin{equation}
\delta(y)=\mathbb{I}\!\left[y \neq \text{original}\right],
\end{equation}
where $y \in \{y_{\text{text}}, y_{\text{image}}, y_{\text{cross}}\}$ denotes the categorical prediction produced by a detection agent. The indicator $\delta(y)=1$ implies that a distortion is detected, whereas $\delta(y)=0$ indicates that the content is classified as original.

Based on $\delta(\cdot)$, the activation indicators for the three detection agents are defined as:
\begin{equation}\label{eq_cascade}
\begin{aligned}
\alpha_{\text{text}}  &= 1, \\
\alpha_{\text{image}} &= 1 - \delta(y_{\text{text}}), \\
\alpha_{\text{cross}} &= \left(1 - \delta(y_{\text{text}})\right)\cdot
                         \left(1 - \delta(y_{\text{image}})\right),
\end{aligned}
\end{equation}
where $\alpha_k \in \{0,1\}$ denotes the activation status of the $k$-th detection agent, with $\alpha_k=1$ indicating that the agent is invoked and $\alpha_k=0$ indicating that it is skipped.
Intuitively, E.q.~\ref{eq_cascade} defines a conditional cascading inference process. AgentM$^3$D starts with textual veracity detection. Only if the textual content is classified as original does the framework proceed to visual veracity detection. Similarly, the cross-modal consistency detection agent is activated only when both the textual and visual contents are deemed original. Once any agent detects a distortion, the cascade terminates and the input is classified as misinformation. An input is regarded as trustworthy only if it passes all activated detection stages.

\subsection{Best-of-N Reasoning with Critique-aware Ranking}
\label{sec:critque_agent}
The hierarchical modality-specific detection framework introduced in the previous section provides a structured and effective decomposition of the multi-modal misinformation detection task. However, the detection agents described so far operate under a one-time forward reasoning paradigm, where each agent produces a single reasoning outcome conditioned on a fixed prompt.
While computationally efficient, one-pass inference is often insufficient to fully exploit the reasoning capacity of modern VLMs. Due to the stochastic nature of VLM inference, single-shot reasoning may lead to unstable predictions and occasional reasoning failures, particularly in ambiguous or challenging cases. As a result, the detection performance of the hierarchical cascade can be suboptimal even when the overall framework design is sound.

To address this limitation, we introduce a test-time scaling mechanism that enhances detection robustness and stimulates VLMs' potentials by encouraging multiple reasoning trajectories during inference. Specifically, we adopt Best-of-$N$ (BoN) reasoning~\cite{sessa2407bond} to generate and evaluate multiple candidate solutions, enabling the detection agents to mitigate stochastic errors and better leverage their intrinsic reasoning capabilities. We adopt BoN as it naturally aligns with the hierarchical cascade of AgentM$^3$D. By selecting the most reliable candidate at each detection stage, BoN yields cleaner intermediate decisions and avoids noise accumulation from aggregating heterogeneous modal evidence~\cite{sessa2407bond}. Besides, its parallel and non-iterative nature further enables seamless integration with our early-stopping and planning mechanisms, allowing efficient and robust test-time inference. The overall process of Best-of-$N$ reasoning with critique-aware ranking is shown in Appendix~\ref{sec:algorithm}

\noindent \textbf{Critique-aware Best-of-$N$ Reasoning.}
For each detection agent $k \in \{\text{text}, \text{image}, \text{cross}\}$, we equip it with a BoN test-time scaling strategy to improve inference robustness. Specifically, under a fixed prompt configuration $\mathcal{P}$, the agent is invoked multiple times in parallel to generate a set of diverse reasoning candidates:
\begin{equation}
\{(y_k^{(n)}, r_k^{(n)})\}_{n=1}^{N} \sim \phi_k(\mathcal{P}).
\end{equation}

Each candidate is first evaluated by a pretrained reward model $\mathcal{R}$ (e.g.,~\cite{yang2024regularizing,liu2025skywork}), which assigns a scalar score based on the predicted label and its associated reasoning:
\begin{equation}
u_k^{(n)} = \mathcal{R}\!\left(y_k^{(n)}, r_k^{(n)}\right),
\end{equation}
where $u_k^{(n)}$ denotes the normalized reward score, with higher values indicating more reliable predictions.

While reward models provide a signal over reasoning quality, they are typically trained on general-purpose QA or instruction-following data and may not be fully aligned with our modality-specific validity requirements of misinformation detection. Specifically, these reward models may fail to identify the factual inconsistencies in these textual or visual claims and only provide the superficial judgment of the semantic coherence between label prediction and its associated reasoning explanation.

To address this limitation, we further introduce modality-specific critique agents for textual and visual detection, which explicitly verify modality-level validity using task-relevant external signals.
Formally, the critique agents are defined as:
\begin{equation}
\begin{aligned}
q_{\text{text}}^{(n)}   &\sim g_{\text{text}}\!\left(
y_{\text{text}}^{(n)}, r_{\text{text}}^{(n)},
\mathcal{T}_{\text{logic}}, \mathcal{P}_{\text{text}}^{\text{crit}}
\right), \\
q_{\text{image}}^{(n)} &\sim g_{\text{image}}\!\left(
y_{\text{image}}^{(n)}, r_{\text{image}}^{(n)},
\mathcal{T}_{\text{forgery}}, \mathcal{P}_{\text{image}}^{\text{crit}}
\right),
\end{aligned}
\end{equation}
where $q_k^{(n)}$ denotes the critique score assigned to the $n$-th candidate. $\mathcal{T}_{\text{logic}}$ and $\mathcal{T}_{\text{forgery}}$ are outputs from the logic consistency checking tool and the image forgery detection tool (see details in Appendix~\ref{sec:toolset}), respectively, while $\mathcal{P}_k^{\text{crit}}$ denotes the critique prompt. Each critique agent evaluates both the predicted label and the corresponding reasoning, providing an orthogonal modality-specific factual validity signal complementary to reward-based evaluation.

Note that, we do not introduce a critique agent for cross-modal consistency detection. This design choice stems from the fact that cross-modal consistency focuses on semantic compatibility between modalities rather than verifying the factual correctness of individual content. Accordingly, the reliability of a cross-modal prediction can be effectively assessed by evaluating the semantic coherence between the predicted label and its accompanying reasoning, for which the general reward model provides a sufficient preference signal. We also empirically investigate the necessity of using a critique agent for cross-modality detection in Sec.~\ref{sec:ablation_study}, which shows degraded performance, likely due to the introduction of redundant or noisy signals.

Finally, we integrate reward-based and critique-based signals into a unified scoring function to guide BoN selection:
\begin{equation}
s_k^{(n)} =
\begin{cases}
u_k^{(n)} + q_k^{(n)}, & k \in \{\text{text}, \text{image}\}, \\
u_k^{(n)},            & k = \text{cross},
\end{cases}
\label{eq:fused_score}
\end{equation}
where $u_k^{(n)}$ and $q_k^{(n)}$ are independently normalized to a common scale prior to fusion. The resulting score $s_k^{(n)}$ enables consistent and fact-aware ranking of reasoning candidates, steering BoN inference toward predictions that better satisfy modality-specific misinformation detection criteria.

\noindent \textbf{Early-Stopping Ranking via Top-$m$ Average Gap.}
Evaluating all $N$ candidates under BoN reasoning can be computationally inefficient, particularly when a clearly dominant solution emerges early in the ranking. To improve test-time efficiency without sacrificing reliability, we introduce an adaptive early-stopping mechanism based on the relative score gap between the top-ranked candidate and its competitors.

Let $\mathbf{s}_k = (s_k^{(1)}, \dots, s_k^{(N)})$ denote the fused scores produced for agent $k$'s reasoning candidates. We sort the scores in descending order to obtain $\mathbf{s}_{k,\downarrow}$, and denote by $\mathbf{s}_{k,\downarrow}^{(1:m)}$ the top-$m$ ranked scores. To quantify the confidence of the leading candidate, we define the Top-$m$ Average Gap as
\begin{equation}
\Delta_m(\mathbf{s}_k) =
s_{k,\downarrow}^{(1)} - \frac{1}{m-1}\sum_{j=2}^{m} s_{k,\downarrow}^{(j)},
\qquad m \ge 2.
\end{equation}
This metric measures how strongly the highest-ranked candidates separate from the remaining top competitors.

The stopping point is determined by identifying the smallest $m$ for which the top-ranked candidates sufficiently outperform the others:
\begin{equation}
m_k^{*} =
\min \Big\{ m \in \{2,\dots,N\} \;\big|\; \Delta_m(\mathbf{s}_k) > \tau \Big\},
\end{equation}
where $\tau$ is a predefined confidence threshold. If no such $m$ satisfies the condition, we set $m_k^{*} = N$, corresponding to full evaluation. The impact of $\tau$ on detection performance and inference efficiency is systematically analyzed in Sec.~\ref{sec:param_impact}.

Finally, the candidate with the highest fused score among the top-$m_k^{*}$ candidates is selected as the output of agent $k$:
\begin{equation}
n_k^{*} = \arg\max_{n \in \{1,\dots,m_k^{*}\}} \; s_{k,\downarrow}^{(n)},
\qquad
(y_k^{*}, r_k^{*}) = (y_k^{(n_k^{*})}, r_k^{(n_k^{*})}).
\end{equation}

The selected pair $(y_k^{*}, r_k^{*})$ serves as the final output of agent $k$ and is subsequently used for hierarchical activation and decision making in E.q.~\ref{eq_cascade}. 

\subsection{Adaptive Test-time Scaling Planning}
\label{sec:planning_agent}

Although critique-aware Best-of-$N$ reasoning improves detection performance, applying it uniformly to all inputs is computationally inefficient and often unnecessary, as many multi-modal samples can be reliably resolved with a single forward pass. This motivates the introduction of an adaptive planning mechanism that selectively allocates test-time computation based on input difficulty.

Specifically, AgentM$^3$D incorporates a lightweight planning agent that determines whether enhanced reasoning should be activated for a given input sample. Since the planning agent is only required to perform a coarse-grained assessment of sample difficulty, rather than fine-grained factual verification, it is implemented as a lightweight, prompt-based decision module without tool invocation. Formally, given a text-image pair $(T, V)$, the planning agent produces a reasoning action $\mathcal{A}$ as
\begin{equation}
\mathcal{A} \;\sim\; \phi_{\mathrm{plan}}\!\left(T,\; V,\; \mathcal{P}_{\mathrm{plan}}\right),
\end{equation}
where $\phi_{\mathrm{plan}}$ denotes the planning agent parameterized by prompt $\mathcal{P}_{\mathrm{plan}}$, and $\mathcal{A}$ specifies the selected inference strategy (i.e., standard single-pass reasoning or enhanced Best-of-$N$ inference) for downstream detection agents.

\subsection{Probabilistic Interpretation of Detection Agent Inference}
From a probabilistic perspective, the inference of each detection agent can be interpreted as approximate posterior reasoning conditioned on both the input and upstream detection outcomes. Specifically, for a detection agent $k \in \{\text{text}, \text{image}, \text{cross}\}$, we model its output as a conditional distribution:
\begin{equation}
p_k(y, r \mid x_k, \mathbf{y}_{<k})
\;\propto\;
\exp\!\Big(
\beta \, S_k(y,r \mid x_k, \mathbf{y}_{<k})
\Big),
\end{equation}
where $x_k$ denotes the agent-specific input (e.g., $T$, $V$, or $(T,V)$), $\mathbf{y}_{<k}$ represents the detection results produced by all preceding agents in the hierarchy. The exponential form induces a preference distribution parameterized by the scoring function. The parameter $\beta>0$ controls how strongly the distribution concentrates on high-scoring candidates, trading off exploitation and exploration.

The scoring function $S_k(\cdot)$ integrates both reward-based and critique-based signals:
\begin{equation}
S_k(y,r \mid x_k, \mathbf{y}_{<k})
=
\mathcal{R}(y,r)
+
q_k(y,r,\mathcal{T}_k),
\end{equation}
where $\mathcal{R}(\cdot)$ denotes the reward model score, and $q_k$ is the critique score produced by the corresponding critique agent based on tool observation $\mathcal{T}_k$.

Under this formulation, BoN corresponds to stochastic sampling from the induced posterior, while critique-aware ranking performs approximate maximum a posteriori estimation over the sampled candidates. The hierarchical activation mechanism further imposes a hard gating prior, ensuring that each agent contributes to the final decision only when activated by preceding inference outcomes.

\begin{table*}[t]
\centering
\caption{Performance comparison on MMDFake and Combined benchmarks. All results are reported in terms of Acc, F1, Recall (Rec), and Precision (Pre). `--' indicates results are not available.}
\label{tab:main_results}
\renewcommand{\arraystretch}{0.8}
\setlength{\tabcolsep}{10pt}
\makebox[\linewidth][c]{%
\begin{tabular}{l l @{\hspace{10pt}} cccc @{\hspace{15pt}} cccc}
\toprule
\textbf{Backbone} & \textbf{Method}
& \multicolumn{4}{c}{\textbf{MMFakeBench}}
& \multicolumn{4}{c}{\textbf{Combined}} \\
\cmidrule(lr){3-6} \cmidrule(lr){7-10}
& & Acc & F1 & Rec & Pre & Acc & F1 & Rec & Pre \\

\midrule

\multirow{6}{*}{Qwen3-VL-4B}
& Standard
& 42.9 & 29.2 & 35.8 & 35.8
& 30.3 & 23.6 & 31.0 & 42.3 \\
& BoN
& 43.7 & 31.1 & 36.4 & 47.9
& 28.2 & 21.9 & 29.5 & 40.5 \\
& T$^2$Agent
& 50.1 & 50.3 & 49.4 & 54.6
& 35.4 & 35.6 & 38.2 & 45.7 \\
& MMD-Agent
& 55.2 & 55.4 & 55.8 & 57.1
& \underline{41.9} & \underline{40.9} & \underline{44.1} & 48.5 \\
& MMD-Agent + BoN
& \underline{57.4} & \underline{57.8} & \underline{58.6} & \textbf{58.5}
& 40.6 & 39.7 & 42.7 & \underline{48.6} \\
& \textbf{AgentM$^3$D (Ours)}
& \textbf{58.1} & \textbf{58.0} & \textbf{60.0} & \underline{57.1}
& \textbf{45.4} & \textbf{45.6} & \textbf{47.3} & \textbf{49.0} \\
\midrule

\multirow{6}{*}{Qwen3-VL-8B}
& Standard
& 46.9 & 37.0 & 39.4 & 59.9
& 33.6 & 28.9 & 36.0 & 40.6 \\
& BoN
& 45.7 & 35.6 & 38.4 & 62.5
& 33.6 & 28.4 & 36.3 & 42.9 \\
& T$^2$Agent
& 54.3 & 54.0 & 52.0 & 61.3
& 36.2 & 36.1 & 38.8 & 45.5 \\
& MMD-Agent
& 59.4 & 60.2 & 60.3 & \underline{62.5}
& \underline{43.3} & \underline{43.5} & \underline{45.2} & \underline{50.5} \\
& MMD-Agent + BoN
& \underline{60.1} & \underline{60.7} & \underline{60.4} & \textbf{62.9}
& 42.3 & 42.6 & 44.3 & 48.7 \\
& \textbf{AgentM$^3$D (Ours)}
& \textbf{62.0} & \textbf{62.6} & \textbf{64.2} & 62.1
& \textbf{48.1} & \textbf{48.3} & \textbf{50.5} & \textbf{52.4} \\
\bottomrule
\end{tabular}
}
\end{table*}

\section{Experiments}
In this section, we conduct experiments to verify the effectiveness of {AgentM$^3$D}, aiming to answer the following research questions (RQs). 
(RQ1) How does AgentM$^3$D perform compared with strong VLM-based baselines and recent agentic methods for multi-modal misinformation detection? 
(RQ2) How well does AgentM$^3$D balance detection performance and inference efficiency, and can it achieve a more favorable accuracy-latency trade-off than existing multi-agent approaches? 
(RQ3) What are the individual contributions of the core components in AgentM$^3$D, including adaptive test-time scaling, critique-aware reasoning, and structured agent coordination, to the overall detection performance? 
(RQ4) How do the hyperparameters in AgentM$^3$D affect the performance and efficiency?
(RQ5) How do the proposed critique agent and adaptive Best-of-$N$ reasoning influence the decision-making process in practice, as demonstrated through case studies of both correctly classified and misclassified instances?

\subsection{Experimental Settings}
\noindent{\textbf{Datasets.}} 
We evaluate AgentM$^3$D on two mixed-source multi-modal misinformation detection (M$^3$D) benchmarks. First, we adopt \textit{MMFakeBench}, which contains over 11,000 image--text pairs spanning four categories: {Real}, {Textual Veracity Distortion (TVD)}, {Visual Veracity Distortion (VVD)}, and {Cross-modal Consistency Distortion (CMM)}, with images originating from both human and machine generation \cite{liu2024mmfakebench}. Following prior agent-based works MMD-Agent and T$^2$Agent \cite{liu2024mmfakebench,t2agent}, we use the same evaluation protocol and sample an identical subset of 1,000 validation instances, balanced as 300 Real, 300 TVD, 100 VVD, and 300 CMM samples.

To further assess generalization under mixed-source settings, we construct a \textit{Combined} benchmark by integrating samples from three complementary datasets: \textit{Mocheg}, a text-centric misinformation dataset with journalist-verified claims \cite{yao2023end}; \textit{Fakeddit-M}, a Reddit-based dataset focusing on manipulated visual content \cite{nakamura2020fakeddit}; and \textit{VERITE}, a real-world benchmark with modality-balanced image--text pairs \cite{papadopoulos2024verite}. From each dataset, we randomly sample 100 real instances and 200 fake instances, resulting in a unified benchmark consisting of 300 Real, 200 TVD, 200 VVD, and 200 CMM samples.  This combined setting enables a systematic evaluation of M$^3$D under diverse content distributions.

\noindent{\textbf{Baselines.}}
We compare AgentM$^3$D with a diverse set of baselines covering three representative paradigms: {VLM-based single-model inference}, {test-time scaling strategies}, and {agentic multi-modal detection frameworks}. These baselines allow us to evaluate both the absolute detection performance and the effectiveness of different reasoning and scaling mechanisms under comparable settings.

\begin{itemize}
\item {Standard}: A vanilla VLM baseline that directly predicts the label from the text-image input via a single forward pass.
\item {Best-of-$N$ (BoN) \cite{irvine2023rewarding}:} A parallel test-time scaling strategy that samples multiple detection outputs from the VLM and selects the final prediction using a reward model.
\item {T$^2$Agent \cite{t2agent}:} An agentic method that formulates multi-modal misinformation detection as a tree search problem and applies Monte Carlo Tree Search to guide multi-step reasoning.
\item {MMD-Agent\cite{liu2024mmfakebench}:} An agentic framework that decomposes multi-modal misinformation detection into multiple task-specific agents with hierarchical reasoning.
\item {MMD-Agent+BoN:} A variant of MMD-Agent where all detection agents are uniformly enhanced with BoN sampling.
\end{itemize}

\noindent{\textbf{Implementation Details.}} 
All experiments are conducted in a zero-shot setting, without any task-specific fine-tuning. We report standard classification metrics, including Accuracy, Precision, Recall, and F1-score. We adopt two vision--language model backbones, \textit{Qwen3-VL-4B-Instruct}\footnote{\url{https://huggingface.co/Qwen/Qwen3-VL-4B-Instruct}} and \textit{Qwen3-VL-8B-Instruct}\footnote{\url{https://huggingface.co/Qwen/Qwen3-VL-8B-Instruct}}, for performance comparison, while all other experiments are conducted using \textit{Qwen3-VL-8B-Instruct}. Following recent agentic frameworks~\cite{xien2025sensenova,bai2026webgym}, we exclusively use models from the Qwen-VL family, as they represent the most powerful open-source VLMs with environment-friendly parameter sizes available at the time of this work~\cite{li2026qwen3}. All experiments are performed on a single NVIDIA RTX~4090 GPU. For Best-of-$N$ reasoning, we set $N=5$, and we also analyze the impact of the hyperparameter $N$ on detection performance in Sec.~\ref{sec:param_N}. All other VLM-related inference parameters follow the default configuration of MMD-Agent~\cite{liu2024mmfakebench}. For all experiments involving Best-of-$N$, we use \textit{GRM-Gemma2-2B-rewardmodel-ft}~\cite{yang2024regularizing} as the reward model for candidate selection. The prompt designs of the proposed agents used in AgentM$^3$D are shown in Appendix~\ref{sec:prompt}.


\subsection{Performance Comparison (RQ1)}
We evaluate the proposed AgentM$^3$D on two M$^3$D benchmarks under a zero-shot setting, and compare it with strong VLM-based and agentic state-of-the-art methods. The quantitative results are summarized in Table~\ref{tab:main_results}. Based on the results, we draw the following observations:

\begin{itemize}
    \item Compared with both VLM-based baselines and existing agentic methods, AgentM$^3$D achieves the strongest performance across nearly all evaluation metrics and benchmarks, demonstrating its effectiveness for zero-shot M$^3$D under diverse content distributions.

    \item Simply augmenting standard VLM inference or existing agentic frameworks with BoN sampling does not consistently yield performance improvements, and performance degradation is frequently observed. This indicates that directly combining BoN with VLM reasoning or agentic system is insufficient to guarantee reliable gains. In contrast, AgentM$^3$D introduces a critique agent to explicitly evaluate and filter candidate predictions, which compensates for the shortcomings of naive BoN usage and enables consistent improvements across different settings.

\end{itemize}

\begin{figure}
    \centering
    \includegraphics[width=.9\linewidth]{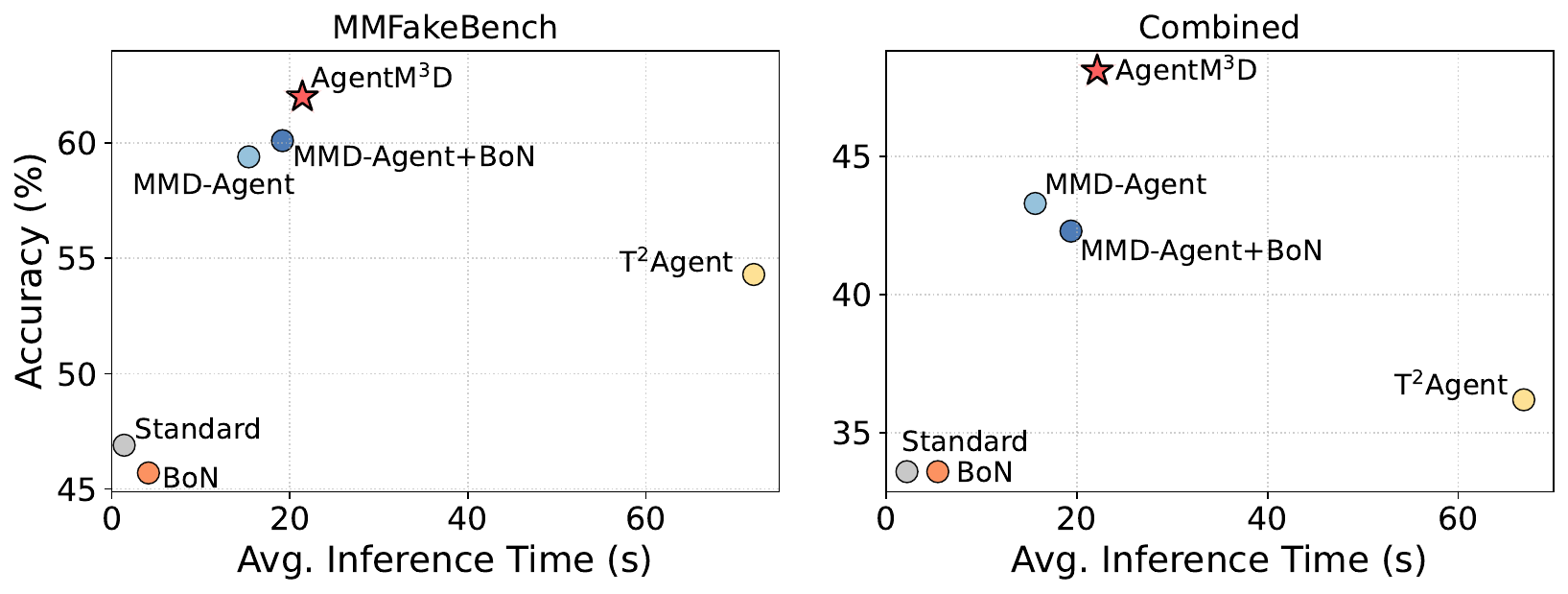}
    \caption{Inference efficiency comparison between AgentM$^3$D and baseline methods.}
    \label{fig:efficiency_tradeoff}
    \vspace{-0.5cm}
\end{figure}


\subsection{Analysis of Inference Efficiency  (RQ2)}
We compare the inference efficiency of AgentM$^3$D with baselines in terms of the accuracy--latency trade-off under a zero-shot setting, as shown in Fig.~\ref{fig:efficiency_tradeoff}. Across both MMFakeBench and the Combined benchmark, methods based on naive test-time scaling (including T$^2$Agent) exhibit substantially higher inference latency than their vanilla counterparts, while delivering limited or inconsistent accuracy gains. In contrast, AgentM$^3$D achieves the highest accuracy with a moderate increase in inference time, positioning it in a more favorable trade-off region than both standard VLM inference and existing agentic baselines. Notably, the proposed planning agent selectively triggers Best-of-$N$ reasoning for only a subset of samples, with 69.1\% of instances on MMFakeBench and 77.2\% on the Combined benchmark being identified as requiring enhanced reasoning. These results indicate that adaptive test-time scaling combined with critique-aware candidate selection enables AgentM$^3$D to improve detection reliability without incurring excessive computational overhead, making it well suited for practical M$^3$D.

\begin{figure}
    \centering
    \includegraphics[width=.9\linewidth]{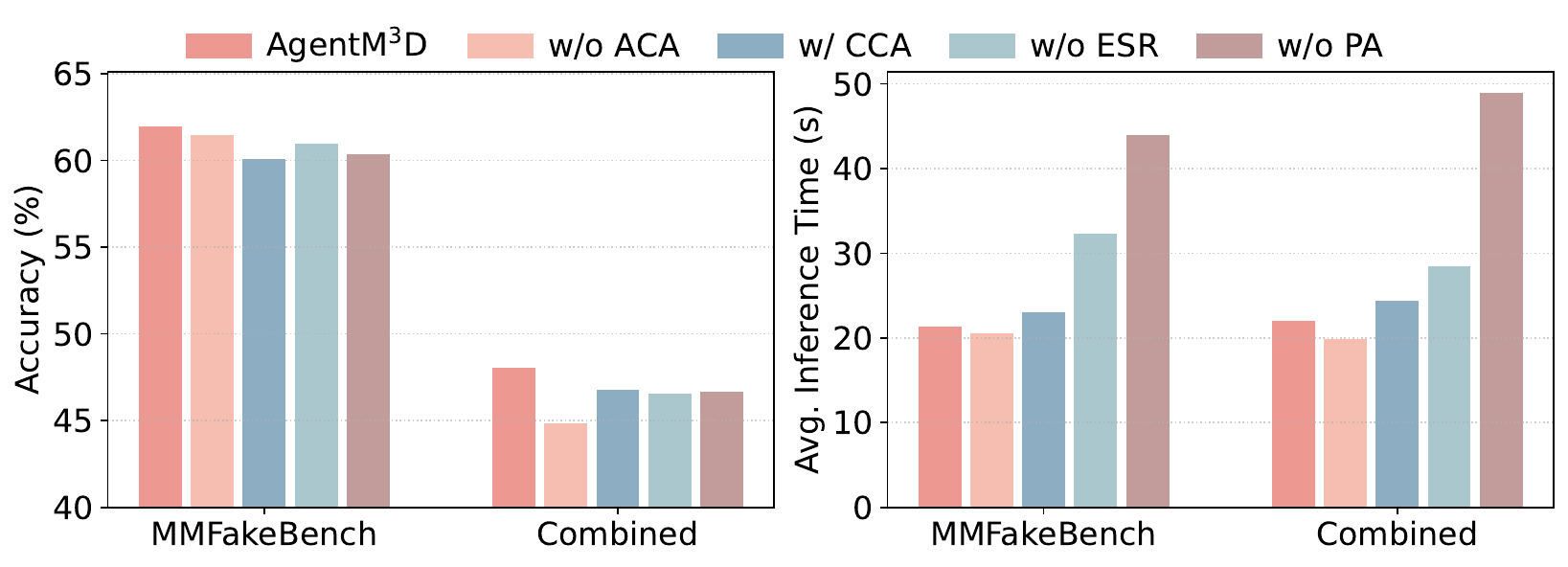}
    \caption{Impact of key components of AgentM$^3$D.}
    \label{fig:ablation_study}
    \vspace{-0.5cm}
\end{figure}

\subsection{Ablation Study (RQ3)}
\label{sec:ablation_study}
We conduct ablation studies on AgentM$^3$D on MMFakeBench and the Combined benchmark to evaluate the contribution of its key components (Fig.~\ref{fig:ablation_study}). Removing early-stopping ranking ({w/o ESR}) significantly increases inference latency with only marginal accuracy changes, demonstrating its effectiveness for efficiency improvement. Disabling all critique agents ({w/o ACA}) consistently degrades accuracy, especially on the Combined benchmark, highlighting the importance of critique-aware assessment for reliable M$^3$D. In contrast, adding an additional cross-modal critique agent ({w/ CCA}) leads to slight performance degradation, suggesting diminishing returns from excessive critique. Finally, removing the planning agent ({w/o PA}) substantially increases inference time without accuracy gains, underscoring the role of planning in efficient test-time reasoning. Overall, these results validate the joint design of planning, critique-aware BoN, and early stopping in AgentM$^3$D.

\begin{figure}
    \centering
    \includegraphics[width=.9\linewidth]{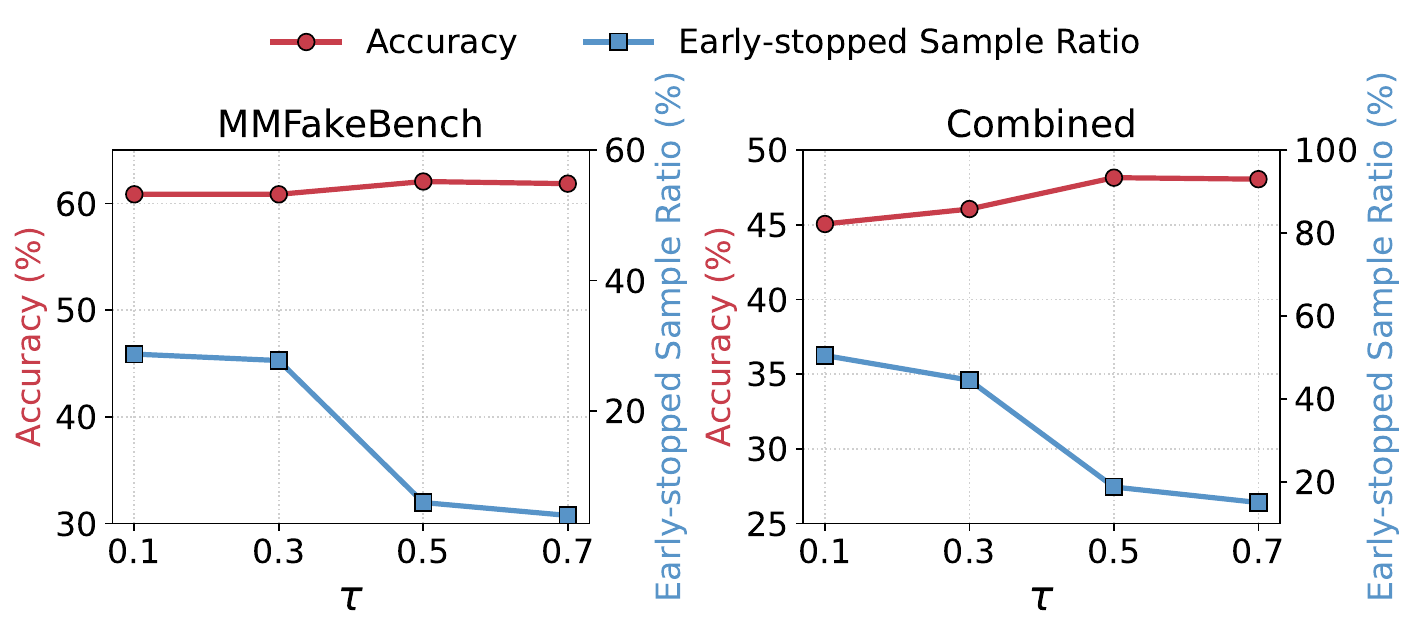}
    \caption{Effect of the early-stopping threshold $\tau$ on detection accuracy and early-stopped sample ratio.}
    \label{fig:param_analysis}
    \vspace{-0.5cm}
\end{figure}

\begin{figure}
    \centering
    \includegraphics[width=0.9\linewidth]{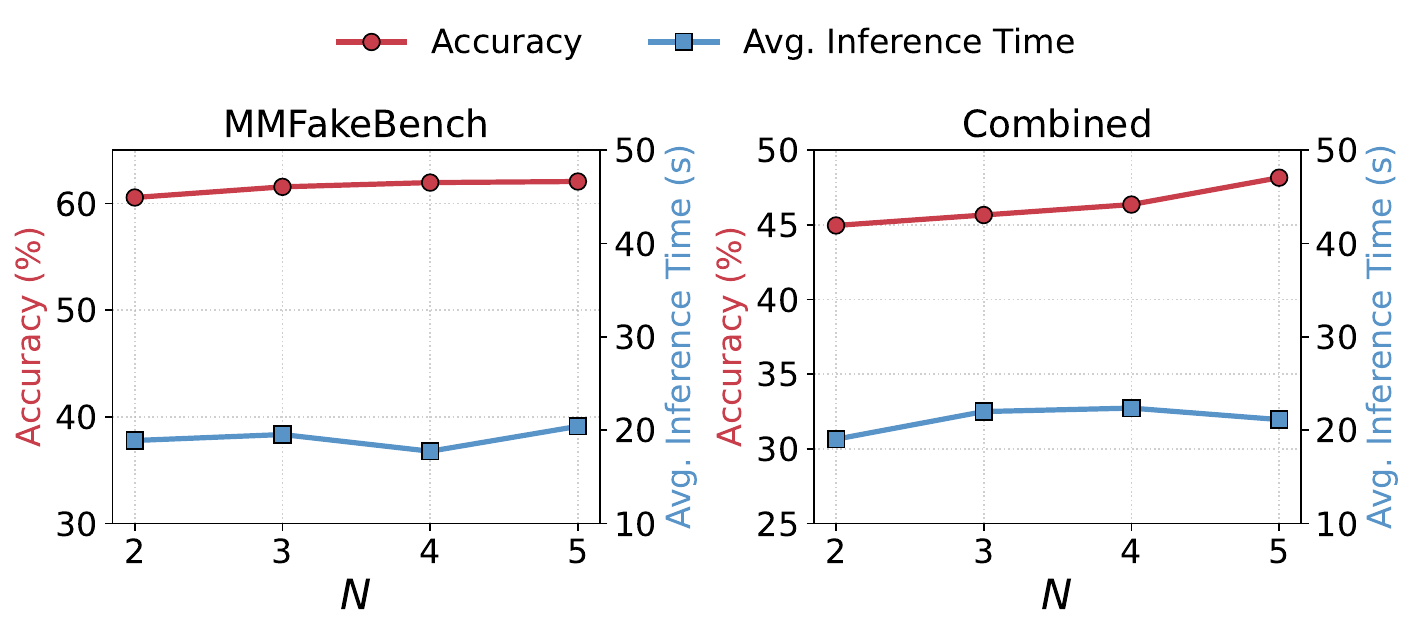}
    \caption{Effect of $N$ in BoN reasoning on detection accuracy and inference time.}
    \label{fig:param_analysis_N}
    \vspace{-0.7cm}
\end{figure}

\subsection{Hyperparameter Analysis (RQ4)}
\subsubsection{Impact of the Hyperparameter $\tau$ in Early-Stopping Ranking}
\label{sec:param_impact}
We analyze the effect of the early-stopping threshold $\tau$ on detection performance and inference behavior. As shown in Fig.~\ref{fig:param_analysis}, increasing $\tau$ reduces the proportion of samples triggering early stopping, resulting in more extensive BoN evaluation. Accuracy improves as $\tau$ increases from 0.1 to 0.5, indicating a balanced regime where low-quality candidates are suppressed without overly restricting candidate exploration, while further increasing $\tau$ to 0.7 yields no additional gains and may introduce ranking noise. Overall, $\tau$ effectively controls the accuracy--efficiency trade-off, with moderate values providing the most favorable balance in practice.

\subsubsection{Impact of $N$ in BoN Reasoning}
\label{sec:param_N}
We conduct a sensitivity analysis on the number of candidates $N$ used in BoN reasoning, as shown in Fig.~\ref{fig:param_analysis_N}. Consistent with prior studies, increasing $N$ generally leads to improved detection accuracy by enabling broader exploration of reasoning trajectories~\cite{sessa2407bond}. Following common practice in agentic methods that adopt BoN reasoning~\cite{li2025metal}, we select $N=5$ as a balanced setting that provides reliable performance gains while remaining computationally affordable in our experimental environment. As BoN reasoning is only activated for a subset of samples by the planning agent, increasing $N$ does not lead to a proportional increase in computation.

\subsection{Case Study (RQ5)}

\begin{figure}
    \centering
    \includegraphics[width=0.7\linewidth]{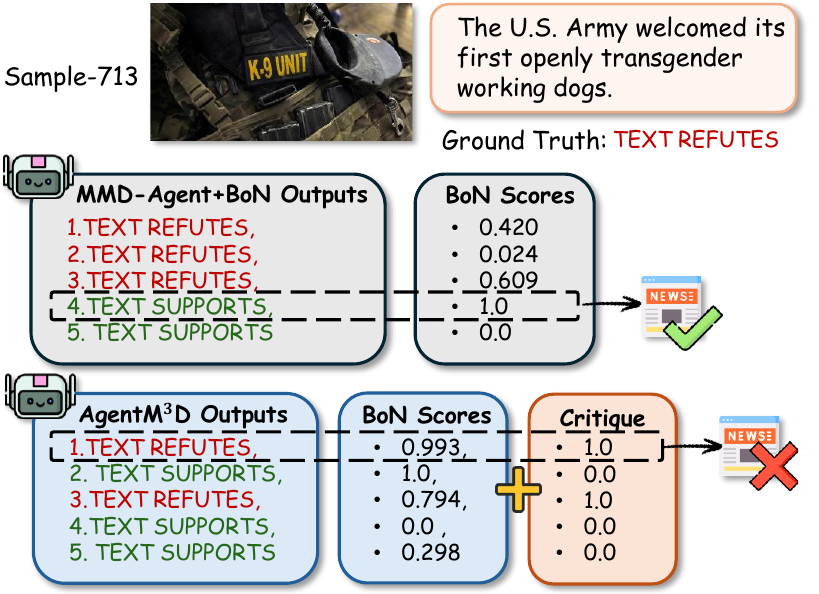}
    \caption{Case Study on Combined Benchmark.}
    \label{fig:case_study}
    \vspace{-.5cm}
\end{figure}

Figure~\ref{fig:case_study} presents a representative case from the Combined benchmark involving a textual veracity distortion. Although both MMD-Agent+BoN and AgentM$^3$D generate multiple reasoning candidates with mixed predictions, their decision mechanisms differ substantially. For MMD-Agent+BoN, the final prediction is dominated by a single high-scoring reasoning path that incorrectly supports the claim, leading to an erroneous decision despite the presence of refuting candidates. Whereas, AgentM$^3$D incorporates critique-aware assessment to down-weight unsupported reasoning paths and emphasize candidates that are both high-confidence and critique-consistent. By aggregating reliable reasoning trajectories rather than relying solely on raw BoN scores, AgentM$^3$D correctly identifies the textual veracity distortion in this example.

\section{Related Work}
\subsection{Multi-Modal Misinformation Detection}
Multi-modal misinformation detection (MMD) has been extensively studied with approaches that either analyze internal inconsistencies across modalities or verify claims using external evidence. Prior work models cross-modal correlations, modality-specific semantics, or visual artifacts to detect textual, visual, and cross-modal distortions \cite{wang2018eann,khattar2019mvae,jiang2025epidemiology,papadopoulos2025similarity,jiang2025towards,wang2026pamas,yin2025graph}, while others incorporate web retrieval, entity matching, or verified repositories for evidence-based verification \cite{papadopoulos2025red,yan2025debunk}. Recent advances further leverage multi-modal large language models (LLMs) and hierarchical reasoning to improve robustness \cite{qi2024sniffer,beigi2024can}. However, most methods assume a single manipulation source and generalize poorly to complex, mixed misinformation. To address this, recent agentic frameworks for zero-shot M$^3$D have emerged. MMD-agent \cite{liu2024mmfakebench} adopts a cascaded multi-agent design that sequentially performs textual, visual, and cross-modal detection, while $\text{T}^2$Agent \cite{t2agent} enhances the agentic detection framework as a search problem using Monte Carlo Tree Search. Despite improved zero-shot capability, these approaches suffer from high test-time cost and loosely coupled reasoning modules. In parallel, TRUST-VL \cite{yan2025trust} targets mixed-source misinformation detection by fine-tuning on large-scale annotated data, which limits its applicability to zero-shot settings considered in this work. In response, we propose AgentM$^3$D, a decision-aware multi-agent framework that enhances zero-shot MMD via adaptive test-time scaling. AgentM$^3$D integrates lightweight planning, critique-aware Best-of-$N$, and adaptive early stopping to selectively allocate computation, enabling robust and cost-efficient inference.

\subsection{LLM-based Multi-Agent Systems}
Recent studies have advocated a paradigm shift from monolithic models toward compound systems composed of multiple specialized components, with multi-agent frameworks emerging as a prominent instantiation of this idea. LLM-based multi-agent systems have been explored across diverse domains, including narrative generation, financial trading, and cooperative problem solving, demonstrating improved reasoning, planning, and decision-making compared to single-agent counterparts \cite{guo2024large,du2023improving,xiao2024tradingagents,plaat2025agentic}. To structure inter-agent collaboration, a variety of organizational topologies have been investigated, such as sequential pipelines, centralized hub-and-spoke designs, hierarchical tree structures, and graph-based interaction patterns \cite{wu2024autogen,qian2024chatdev}. More recent frameworks further emphasize layered or role-specialized organizations to coordinate agent behaviors effectively. Despite their success, existing multi-agent systems have not yet provided a satisfactory agentic solution to the challenges of zero-shot M$^3$D, such as mitigating cross-modal error propagation and regulating test-time computation in a principled manner. In this work, we propose AgentM$^3$D, a multi-agent framework tailored to MMD, addressing the limitations of existing multi-agent systems in mixed-source and zero-shot settings.

\section{Conclusion}
In this work, we present AgentM$^3$D, a multi-agent framework for mixed-source multi-modal misinformation detection that integrates both reliability and test-time efficiency. By integrating adaptive test-time scaling with critique-aware Best-of-$N$ reasoning, AgentM$^3$D selectively allocates computation based on confidence signals, avoiding unnecessary candidate exploration while maintaining robust detection performance. Extensive experiments on MMFakeBench and a newly constructed mixed-source benchmark demonstrate that AgentM$^3$D consistently outperforms strong VLM-based and agentic baselines in a zero-shot setting, while achieving a favorable accuracy--efficiency trade-off. Overall, AgentM$^3$D provides a practical and extensible foundation for deploying reliable multi-modal misinformation detection systems under realistic computational constraints.

\bibliographystyle{ACM-Reference-Format}
\balance
\bibliography{sample-base}

@inproceedings{wang2018eann,
  title={Eann: Event adversarial neural networks for multi-modal fake news detection},
  author={Wang, Yaqing and Ma, Fenglong and Jin, Zhiwei and Yuan, Ye and Xun, Guangxu and Jha, Kishlay and Su, Lu and Gao, Jing},
  booktitle={Proceedings of the 24th acm sigkdd international conference on knowledge discovery \& data mining},
  pages={849--857},
  year={2018}
}

@inproceedings{jiang2025epidemiology,
  title={Epidemiology-informed network for robust rumor detection},
  author={Jiang, Wei and Chen, Tong and Gao, Xinyi and Zhang, Wentao and Cui, Lizhen and Yin, Hongzhi},
  booktitle={Proceedings of the ACM on Web Conference 2025},
  pages={3618--3627},
  year={2025}
}

@article{jiang2025towards,
  title={Towards Propagation-aware Representation Learning for Supervised Social Media Graph Analytics},
  author={Jiang, Wei and Chen, Tong and Yuan, Wei and Zhao, Xiangyu and Nguyen, Quoc Viet Hung and Yin, Hongzhi},
  journal={arXiv preprint arXiv:2509.01124},
  year={2025}
}

@inproceedings{qi2024sniffer,
  title={Sniffer: Multimodal large language model for explainable out-of-context misinformation detection},
  author={Qi, Peng and Yan, Zehong and Hsu, Wynne and Lee, Mong Li},
  booktitle={Proceedings of the IEEE/CVF conference on computer vision and pattern recognition},
  pages={13052--13062},
  year={2024}
}

@inproceedings{yan2025trust,
  title={Trust-vl: An explainable news assistant for general multimodal misinformation detection},
  author={Yan, Zehong and Qi, Peng and Hsu, Wynne and Lee, Mong-Li},
  booktitle={Proceedings of the 2025 Conference on Empirical Methods in Natural Language Processing},
  pages={5588--5604},
  year={2025}
}

@inproceedings{khattar2019mvae,
  title={Mvae: Multimodal variational autoencoder for fake news detection},
  author={Khattar, Dhruv and Goud, Jaipal Singh and Gupta, Manish and Varma, Vasudeva},
  booktitle={The world wide web conference},
  pages={2915--2921},
  year={2019}
}

@article{beigi2024can,
  title={Can LLMs Improve Multimodal Fact-Checking by Asking Relevant Questions?},
  author={Beigi, Alimohammad and Jiang, Bohan and Li, Dawei and Tan, Zhen and Shaeri, Pouya and Kumarage, Tharindu and Bhattacharjee, Amrita and Liu, Huan},
  journal={arXiv preprint arXiv:2410.04616},
  year={2024}
}

@article{papadopoulos2025red,
  title={Red-dot: Multimodal fact-checking via relevant evidence detection},
  author={Papadopoulos, Stefanos-Iordanis and Koutlis, Christos and Papadopoulos, Symeon and Petrantonakis, Panagiotis C},
  journal={IEEE Transactions on Computational Social Systems},
  year={2025},
  publisher={IEEE}
}

@inproceedings{papadopoulos2025similarity,
  title={Similarity over Factuality: Are we making progress on multimodal out-of-context misinformation detection?},
  author={Papadopoulos, Stefanos-Iordanis and Koutlis, Christos and Papadopoulos, Symeon and Petrantonakis, Panagiotis C},
  booktitle={2025 IEEE/CVF Winter Conference on Applications of Computer Vision (WACV)},
  pages={5041--5050},
  year={2025},
  organization={IEEE}
}

@article{yan2025debunk,
  title={Debunk and Infer: Multimodal Fake News Detection via Diffusion-Generated Evidence and LLM Reasoning},
  author={Yan, Kaiying and Liu, Moyang and Liu, Yukun and Fu, Ruibo and Wen, Zhengqi and Tao, Jianhua and Liu, Xuefei},
  journal={arXiv preprint arXiv:2506.21557},
  year={2025}
}

@article{liu2024mmfakebench,
  title={Mmfakebench: A mixed-source multimodal misinformation detection benchmark for lvlms},
  author={Liu, Xuannan and Li, Zekun and Li, Peipei and Huang, Huaibo and Xia, Shuhan and Cui, Xing and Huang, Linzhi and Deng, Weihong and He, Zhaofeng},
  journal={arXiv preprint arXiv:2406.08772},
  year={2024}
}

@article{t2agent,
  title={T\^{} 2Agent A Tool-augmented Multimodal Misinformation Detection Agent with Monte Carlo Tree Search},
  author={Cui, Xing and Zou, Yueying and Li, Zekun and Li, Peipei and Xu, Xinyuan and Liu, Xuannan and Huang, Huaibo},
  journal={arXiv preprint arXiv:2505.19768},
  year={2025}
}

@inproceedings{qian2024chatdev,
  title={Chatdev: Communicative agents for software development},
  author={Qian, Chen and Liu, Wei and Liu, Hongzhang and Chen, Nuo and Dang, Yufan and Li, Jiahao and Yang, Cheng and Chen, Weize and Su, Yusheng and Cong, Xin and others},
  booktitle={Proceedings of the 62nd Annual Meeting of the Association for Computational Linguistics (Volume 1: Long Papers)},
  pages={15174--15186},
  year={2024}
}

@inproceedings{wu2024autogen,
  title={Autogen: Enabling next-gen LLM applications via multi-agent conversations},
  author={Wu, Qingyun and Bansal, Gagan and Zhang, Jieyu and Wu, Yiran and Li, Beibin and Zhu, Erkang and Jiang, Li and Zhang, Xiaoyun and Zhang, Shaokun and Liu, Jiale and others},
  booktitle={First Conference on Language Modeling},
  year={2024}
}

@inproceedings{du2023improving,
  title={Improving factuality and reasoning in language models through multiagent debate},
  author={Du, Yilun and Li, Shuang and Torralba, Antonio and Tenenbaum, Joshua B and Mordatch, Igor},
  booktitle={Forty-first International Conference on Machine Learning},
  year={2023}
}

@article{guo2024large,
  title={Large language model based multi-agents: A survey of progress and challenges},
  author={Guo, Taicheng and Chen, Xiuying and Wang, Yaqi and Chang, Ruidi and Pei, Shichao and Chawla, Nitesh V and Wiest, Olaf and Zhang, Xiangliang},
  journal={arXiv preprint arXiv:2402.01680},
  year={2024}
}

@article{xiao2024tradingagents,
  title={TradingAgents: Multi-agents LLM financial trading framework},
  author={Xiao, Yijia and Sun, Edward and Luo, Di and Wang, Wei},
  journal={arXiv preprint arXiv:2412.20138},
  year={2024}
}

@article{plaat2025agentic,
  title={Agentic large language models, a survey},
  author={Plaat, Aske and van Duijn, Max and van Stein, Niki and Preuss, Mike and van der Putten, Peter and Batenburg, Kees Joost},
  journal={arXiv preprint arXiv:2503.23037},
  year={2025}
}

@inproceedings{yao2023end,
  title={End-to-end multimodal fact-checking and explanation generation: A challenging dataset and models},
  author={Yao, Barry Menglong and Shah, Aditya and Sun, Lichao and Cho, Jin-Hee and Huang, Lifu},
  booktitle={Proceedings of the 46th International ACM SIGIR Conference on Research and Development in Information Retrieval},
  pages={2733--2743},
  year={2023}
}

@inproceedings{nakamura2020fakeddit,
  title={Fakeddit: A new multimodal benchmark dataset for fine-grained fake news detection},
  author={Nakamura, Kai and Levy, Sharon and Wang, William Yang},
  booktitle={Proceedings of the twelfth language resources and evaluation conference},
  pages={6149--6157},
  year={2020}
}

@article{papadopoulos2024verite,
  title={Verite: a robust benchmark for multimodal misinformation detection accounting for unimodal bias},
  author={Papadopoulos, Stefanos-Iordanis and Koutlis, Christos and Papadopoulos, Symeon and Petrantonakis, Panagiotis C},
  journal={International Journal of Multimedia Information Retrieval},
  volume={13},
  number={1},
  pages={4},
  year={2024},
  publisher={Springer}
}

@inproceedings{yang2024regularizing,
  title={Regularizing Hidden States Enables Learning Generalizable Reward Model for LLMs},
  author={Yang, Rui and Ding, Ruomeng and Lin, Yong and Zhang, Huan and Zhang, Tong},
  booktitle={Advances in Neural Information Processing Systems},
  year={2024}
}

@article{liu2025skywork,
  title={Skywork-Reward-V2: Scaling Preference Data Curation via Human-AI Synergy},
  author = {Liu, Chris Yuhao and Zeng, Liang and Xiao, Yuzhen and He, Jujie and Liu, Jiacai and Wang, Chaojie and Yan, Rui and Shen, Wei and Zhang, Fuxiang and Xu, Jiacheng and Liu, Yang and Zhou, Yahui},
  journal={arXiv preprint arXiv:2507.01352},
  year={2025}
}

@article{irvine2023rewarding,
  title={Rewarding chatbots for real-world engagement with millions of users},
  author={Irvine, Robert and Boubert, Douglas and Raina, Vyas and Liusie, Adian and Zhu, Ziyi and Mudupalli, Vineet and Korshuk, Aliaksei and Liu, Zongyi and Cremer, Fritz and Assassi, Valentin and others},
  journal={arXiv preprint arXiv:2303.06135},
  year={2023}
}

@article{li2025metal,
  title={Metal: A multi-agent framework for chart generation with test-time scaling},
  author={Li, Bingxuan and Wang, Yiwei and Gu, Jiuxiang and Chang, Kai-Wei and Peng, Nanyun},
  journal={arXiv preprint arXiv:2502.17651},
  year={2025}
}

@book{o2019misinformation,
  title={The misinformation age: How false beliefs spread},
  author={O'Connor, Cailin and Weatherall, James Owen},
  year={2019},
  publisher={Yale University Press}
}

@article{singh2020first,
  title={A first look at COVID-19 information and misinformation sharing on Twitter},
  author={Singh, Lisa and Bansal, Shweta and Bode, Leticia and Budak, Ceren and Chi, Guangqing and Kawintiranon, Kornraphop and Padden, Colton and Vanarsdall, Rebecca and Vraga, Emily and Wang, Yanchen},
  journal={arXiv preprint arXiv:2003.13907},
  year={2020}
}

@article{pennycook2019fighting,
  title={Fighting misinformation on social media using crowdsourced judgments of news source quality},
  author={Pennycook, Gordon and Rand, David G},
  journal={Proceedings of the National Academy of Sciences},
  volume={116},
  number={7},
  pages={2521--2526},
  year={2019},
  publisher={National Academy of Sciences}
}

@inproceedings{lin2023zero,
  title={Zero-shot rumor detection with propagation structure via prompt learning},
  author={Lin, Hongzhan and Yi, Pengyao and Ma, Jing and Jiang, Haiyun and Luo, Ziyang and Shi, Shuming and Liu, Ruifang},
  booktitle={Proceedings of the AAAI Conference on Artificial Intelligence},
  volume={37},
  number={4},
  pages={5213--5221},
  year={2023}
}

@article{abdali2024multi,
  title={Multi-modal misinformation detection: Approaches, challenges and opportunities},
  author={Abdali, Sara and Shaham, Sina and Krishnamachari, Bhaskar},
  journal={ACM Computing Surveys},
  volume={57},
  number={3},
  pages={1--29},
  year={2024},
  publisher={ACM New York, NY}
}

@inproceedings{shang2024mmadapt,
  title={MMAdapt: A knowledge-guided multi-source multi-class domain adaptive framework for early health misinformation detection},
  author={Shang, Lanyu and Zhang, Yang and Chen, Bozhang and Zong, Ruohan and Yue, Zhenrui and Zeng, Huimin and Wei, Na and Wang, Dong},
  booktitle={Proceedings of the ACM Web Conference 2024},
  pages={4653--4663},
  year={2024}
}

@article{zhou2024finefake,
  title={Finefake: A knowledge-enriched dataset for fine-grained multi-domain fake news detection},
  author={Zhou, Ziyi and Zhang, Xiaoming and Zhang, Litian and Liu, Jiacheng and Wang, Senzhang and Liu, Zheng and Zhang, Xi and Li, Chaozhuo and Yu, Philip S},
  journal={arXiv preprint arXiv:2404.01336},
  year={2024}
}

@article{shen2024small,
  title={Small llms are weak tool learners: A multi-llm agent},
  author={Shen, Weizhou and Li, Chenliang and Chen, Hongzhan and Yan, Ming and Quan, Xiaojun and Chen, Hehong and Zhang, Ji and Huang, Fei},
  journal={arXiv preprint arXiv:2401.07324},
  year={2024}
}

@article{zhang2025survey,
  title={A Survey on Test-Time Scaling in Large Language Models: What, How, Where, and How Well?},
  author={Zhang, Qiyuan and Lyu, Fuyuan and Sun, Zexu and Wang, Lei and Zhang, Weixu and Hua, Wenyue and Wu, Haolun and Guo, Zhihan and Wang, Yufei and Muennighoff, Niklas and others},
  journal={arXiv preprint arXiv:2503.24235},
  year={2025}
}

@article{zhang2024vision,
  title={Vision-language models for vision tasks: A survey},
  author={Zhang, Jingyi and Huang, Jiaxing and Jin, Sheng and Lu, Shijian},
  journal={IEEE transactions on pattern analysis and machine intelligence},
  volume={46},
  number={8},
  pages={5625--5644},
  year={2024},
  publisher={IEEE}
}

@article{chen2024expanding,
  title={Expanding performance boundaries of open-source multimodal models with model, data, and test-time scaling},
  author={Chen, Zhe and Wang, Weiyun and Cao, Yue and Liu, Yangzhou and Gao, Zhangwei and Cui, Erfei and Zhu, Jinguo and Ye, Shenglong and Tian, Hao and Liu, Zhaoyang and others},
  journal={arXiv preprint arXiv:2412.05271},
  year={2024}
}

@article{sessa2407bond,
  title={Bond: Aligning llms with best-of-n distillation, 2024},
  author={Sessa, Pier Giuseppe and Dadashi, Robert and Hussenot, L{\'e}onard and Ferret, Johan and Vieillard, Nino and Ram{\'e}, Alexandre and Shariari, Bobak and Perrin, Sarah and Friesen, Abe and Cideron, Geoffrey and others},
  journal={URL https://arxiv. org/abs/2407.14622}
}

@article{zhong2023patchcraft,
  title={Patchcraft: Exploring texture patch for efficient ai-generated image detection},
  author={Zhong, Nan and Xu, Yiran and Li, Sheng and Qian, Zhenxing and Zhang, Xinpeng},
  journal={arXiv preprint arXiv:2311.12397},
  year={2023}
}

@article{xien2025sensenova,
  title={SenseNova-MARS: Empowering Multimodal Agentic Reasoning and Search via Reinforcement Learning},
  author={Xien Chng, Yong and Hu, Tao and Tong, Wenwen and Li, Xueheng and Chen, Jiandong and Yu, Haojia and Lu, Jiefan and Guo, Hewei and Deng, Hanming and Xie, Chengjun and others},
  journal={arXiv e-prints},
  pages={arXiv--2512},
  year={2025}
}

@article{bai2026webgym,
  title={WebGym: Scaling Training Environments for Visual Web Agents with Realistic Tasks},
  author={Bai, Hao and Taymanov, Alexey and Zhang, Tong and Kumar, Aviral and Whitehead, Spencer},
  journal={arXiv preprint arXiv:2601.02439},
  year={2026}
}

@article{li2026qwen3,
  title={Qwen3-VL-Embedding and Qwen3-VL-Reranker: A Unified Framework for State-of-the-Art Multimodal Retrieval and Ranking},
  author={Li, Mingxin and Zhang, Yanzhao and Long, Dingkun and Chen, Keqin and Song, Sibo and Bai, Shuai and Yang, Zhibo and Xie, Pengjun and Yang, An and Liu, Dayiheng and others},
  journal={arXiv preprint arXiv:2601.04720},
  year={2026}
}

@article{wang2026pamas,
  title={PAMAS: Self-Adaptive Multi-Agent System with Perspective Aggregation for Misinformation Detection},
  author={Wang, Zongwei and Gao, Min and Yu, Junliang and Chen, Tong and Lin, Chenghua},
  journal={arXiv preprint arXiv:2602.03158},
  year={2026}
}

@inproceedings{yin2025graph,
  title={Graph with Sequence: Broad-Range Semantic Modeling for Fake News Detection},
  author={Yin, Junwei and Gao, Min and Shu, Kai and Li, Wentao and Huang, Yinqiu and Wang, Zongwei},
  booktitle={Proceedings of the ACM on Web Conference 2025},
  pages={2838--2849},
  year={2025}
}

\clearpage

\appendix
\section{Appendix}
\subsection{Detailed Prompts of the Proposed Agents}
\label{sec:prompt}
\begin{tcolorbox}[
enhanced, colback=white!5, colframe=gray!20!black,
width=\linewidth, arc=1mm, auto outer arc,
title={Planning Agent}, breakable, boxsep=0mm,
top=2mm, bottom=2mm, fontupper=\linespread{0.95}\selectfont
]
Given a multi-modal misinformation sample, it contains both a news caption and a news image.

News caption is: \textcolor{blue}{$<$news caption$>$}

Your task is to decide whether this sample should be handled with standard reasoning or escalated to a stronger reasoning level using Best-of-N (BON) in later stages. Best-of-N (BON) refers to sampling multiple independent detection responses with the same prompt and selecting the most reliable one.

Analyze the given news caption and image from the following aspects:

- Whether the relationship between the caption and the image is clearly consistent, clearly inconsistent, or ambiguous.

- Whether the caption makes claims that require explicit and concrete visual evidence.

- Whether the image content alone is sufficient to verify those claims.
Based on the analysis, choose ONE action from the following options:

1. [BON level-0]: No Best-of-N scaling is needed.

2. [BON level-1]: Use Best-of-N scaling.

Return ONLY the action in the exact form: [BON level-n]
\end{tcolorbox}

\begin{tcolorbox}[
enhanced, colback=white!5, colframe=gray!20!black,
width=\linewidth, arc=1mm, auto outer arc,
title={Textual Critique Agent}, breakable, boxsep=0mm,
top=2mm, bottom=2mm, fontupper=\linespread{0.95}\selectfont
]
Given a news caption, news caption is:

\textcolor{blue}{$<$news caption$>$}

Your task is to assign a single score between 0 and 1 indicating how convincing the DETECTION RESULT is, based ONLY on its own reasoning.

Detection Result:

\textcolor{red}{$<$detection result$>$}

Result from logical consistency checking tool:

\textcolor{red}{$<$logical consistency checking result$>$}

Score the detection result from 0 to 1:

- 1.0: fully convincing, logically sound, no over-inference.

- 0.5: partially convincing, some logical gaps or weak support.

- 0.0: unconvincing, logical  contradiction or strong over-inference.

Output ONLY the score as a number between 0 and 1.
\end{tcolorbox}

\begin{tcolorbox}[
enhanced, colback=white!5, colframe=gray!20!black,
width=\linewidth, arc=1mm, auto outer arc,
title={Visual Critique Agent}, breakable, boxsep=0mm,
top=2mm, bottom=2mm, fontupper=\linespread{0.95}\selectfont
]
According to the given news image, your task is to assign a single score between 0 and 1 indicating how convincing the DETECTION RESULT is, based ONLY on its own reasoning.

Detection Result:

\textcolor{red}{$<$detection result$>$}

Result from image forgery detection tool:

\textcolor{red}{$<$image forgery detection result$>$}

Score the detection result from 0 to 1:

- 1.0: The detection result is 
strongly supported by the image forensic result, with no apparent logical gaps or over-interpretation.

- 0.5: The detection result is partially supported by the image forensic result, but contains uncertainty, weak evidence, or mild over-interpretation.

- 0.0: The detection result is not supported by the image forensic result, or shows clear logical inconsistency or strong over-interpretation.

Output ONLY the score as a number between 0 and 1.
\end{tcolorbox}

\subsection{Algorithm}
\label{sec:algorithm}
The algorithm of overall process of Best-of-$N$ reasoning with critique-aware ranking is shown in Algorithm~\ref{alg:critique-aware_BoN}.
\begin{algorithm}
\caption{Best-of-$N$ reasoning with critique-aware ranking.}
\label{alg:critique-aware_BoN}
\DontPrintSemicolon
\SetKwComment{tcp}{// }{}
\KwIn{Agent $k$ input $\mathcal{X}_k$, candidate budget $N$, threshold $\tau$}
\KwOut{Selected prediction $y^*$ and reasoning $r^*$}

\tcp{Generate BoN candidates}
Sample $N$ reasoning candidates
$\{(y^{(n)}, r^{(n)})\}_{n=1}^{N} \sim \phi_k(\mathcal{X}_k)$\;

\tcp{Reward and critique scoring}
\For{$n \leftarrow 1$ \KwTo $N$}{
    $u^{(n)} \leftarrow \mathcal{R}(y^{(n)}, r^{(n)})$\;
    \If{$k \in \{\text{text},\text{image}\}$}{
        $q^{(n)} \sim g_k(y^{(n)}, r^{(n)}, \mathcal{T}_k)$\;
        $s^{(n)} \leftarrow u^{(n)} + q^{(n)}$\;
    }
    \Else{
        $s^{(n)} \leftarrow u^{(n)}$\;
    }
}

\tcp{Early-stopping ranking via Top-$m$ Average Gap}
Sort $\{s^{(n)}\}_{n=1}^{N}$ in descending order to obtain
$\{s_{\downarrow}^{(j)}\}_{j=1}^{N}$\;

\For{$m \leftarrow 2$ \KwTo $N$}{
    $\Delta_m \leftarrow s_{\downarrow}^{(1)} -
    \frac{1}{m-1}\sum_{j=2}^{m} s_{\downarrow}^{(j)}$\;
    \If{$\Delta_m > \tau$}{
        \textbf{break}\;
    }
}

\tcp{Select best candidate among top-$m$}
$n^* \leftarrow \arg\max_{n \in \{1,\dots,m\}} s_{\downarrow}^{(n)}$\;
\Return $(y^{(n^*)}, r^{(n^*)})$\;

\end{algorithm}

\subsection{Details of Toolset}
\label{sec:toolset}
To support modality-specific verification in AgentM$^3$D, we employ a lightweight and extensible toolset that provides complementary external signals for misinformation detection.

\noindent \textbf{Web Searching.}
We use Wikipedia\footnote{\url{https://www.wikipedia.org/}} to retrieve encyclopedic knowledge for entities mentioned in news text, which serves as external evidence for textual veracity assessment.

\noindent \textbf{Text Logic Consistency Checking.}
We adopt Qwen3-4B-Instruct-2507\footnote{\url{https://huggingface.co/Qwen/Qwen3-4B-Instruct-2507}} as a logic consistency checking tool to analyze whether the textual claim exhibits logical contradictions or implausible reasoning.

\noindent \textbf{Image Forgery Detection.}
A well-trained image forgery detection model~\cite{zhong2023patchcraft} is employed to identify potential digital manipulations in visual content, providing an objective signal for visual authenticity verification.

\noindent \textbf{Image Analysis.}
We use Qwen3-VL-8B-Instruct\footnote{\url{https://huggingface.co/Qwen/Qwen3-VL-8B-Instruct}} to extract grounded descriptions of visual content, which are used for visual veracity assessment and cross-modal consistency checking.

\end{document}